\documentclass[aps,prx,reprint,twocolumn]{revtex4-1}
\usepackage{amsmath,amssymb}
\usepackage{subfigure}
\usepackage{graphicx}
\usepackage{bm}
\usepackage{color}
\usepackage{times}
\usepackage{siunitx}
\usepackage{mathtools}

\begin{document}
\title{Topological Dirac Nodal-net Fermions in AlB$_2$-type TiB$_2$ and ZrB$_2$}
\author{Xing Feng$^{1*}$}
\author{Changming Yue$^2$}
\thanks{X. Feng and C.M. Yue contributed equally to this work}
\author{Zhida Song$^2$}
\author{QuanSheng Wu$^3$}
\email[]{wuq@phys.ethz.ch}
\author{Bin Wen$^1$}
\email[]{wenbin@ysu.edu.cn}

\affiliation{$^1$State Key Laboratory of Metastable Materials Science and Technology, Yanshan University, Qinhuangdao 066004, China}
\affiliation{$^2$Beijing National Laboratory for Condensed Matter Physics, and Institute of Physics, Chinese Academy of Science, Beijing 100190, China}
\affiliation{$^3$Theoretical Physics and Station Q Zurich, ETH Zurich, 8093 Zurich, Switzerland}
\date{\today}

\begin{abstract}
Based on first-principles calculations and effective model analysis, a Dirac nodal-net semimetal state is recognized in AlB$_2$-type TiB$_2$ and ZrB$_2$ when spin-orbit coupling (SOC) is ignored. Taking TiB$_2$ as an example, there are several topological excitations in this nodal-net structure including triple point, nexus, and nodal link, which are protected by coexistence of spatial-inversion symmetry and time reversal symmetry. This nodal-net state is remarkably different from that of IrF$_4$, which requires sublattice chiral symmetry. In addition, linearly and quadratically dispersed two-dimensional surface Dirac points are identified as having emerged on the B-terminated and Ti-terminated (001) surfaces of TiB$_2$ respectively, which are analogous to those of monolayer and bilayer graphene.
\end{abstract}

\maketitle

\section{Introduction}

In the band theory of solids, a nodal point is a point of contact between conduction and valence bands. One way to classify nodal points is based on their dimensionality~\cite{Chiu2016Classification, bzduvsek2016nodal,Kobayashi2017Crossing, PhysRevB.84.235126}, which can be divided into three categories. The first is zero-dimensional (0D) nodal points including Weyl points~\cite{PhysRevB.83.205101, PhysRevX.5.011029, Huang2015, Xu613, PhysRevX.5.031013,soluyanov2015type}, Dirac points~\cite{novoselov2005two,PhysRevB.85.195320,Liu864,PhysRevLett.108.140405}, triple points~\cite{PhysRevX.6.031003, PhysRevB.93.241202, Chang2016}, and other higher degeneracy nodal points~\cite{Bradlynaaf5037}. The second is one-dimensional (1D) nodal line systems which include nodal ring~\cite{PhysRevLett.117.016602,PhysRevB.93.121113}, nodal chain~\cite{bzduvsek2016nodal,Yu2017From} and nodal net~\cite{bzduvsek2016nodal}. The third is two-dimensional (2D) nodal surfaces~\cite{PhysRevB.93.085427}. A semimetal with such nodal points is called a topological semimetal (TS), such as Weyl semimetal (WSM), Dirac semimetal (DSM), Dirac nodal line semimetal (DNLSM), nodal chain semimetal, \emph{et al}.

The 0D nodal point system has been extensively studied during the past decade, due to its non-trivial topological properties and its exotic transport properties. For example, Dirac semimetal has been predicted to be a good candidate for quantum devices because of its massless Dirac fermion's properties~\cite{novoselov2005two,zhang2005experimental,PhysRevB.85.195320,Liu864},  Weyl semimetals have chiral anomaly which leads to a chiral magnetic effect~\cite{PhysRevLett.109.181602,PhysRevB.94.214306} \emph{i.e.} an electric current parallel to an external magnetic field~\cite{PhysRevLett.111.027201,PhysRevD.78.074033,Ali2014,PhysRevB.86.115133,Hosur2013Recent,PhysRevX.5.031023}, and a triple point metal could have topological Lifshitz transitions~\cite{PhysRevX.6.031003}.

Investigations of 1D nodal line systems have been developing recent years, due to their great potential in diverse developments in materials science. A nodal line system has a non-trivial $\pi$ Berry phase around the nodal line which would shift the Landau level index by 1/2~\cite{PhysRevB.77.245413,Yan2017}, and leads to drumhead surface states (SSs)~\cite{Fang2016Topological,Yu2017Topological}. There are many types of nodal line systems, such as single and multiple Dirac nodal line (DNL) systems in the absence of spin-orbit coupling (SOC) \cite{PhysRevB.84.235126,PhysRevB.93.121113,hirayama2017topological,cheng2017body,PhysRevB.93.085427,PhysRevLett.115.036806,PhysRevB.92.081201}, nexus \cite{Heikkila2015Nexus,Hyart2016Momentum,PhysRevX.6.031003}, nodal chain with SOC~\cite{bzduvsek2016nodal}, and other crossing nodal lines \cite{Yu2017From,PhysRevB.92.045108,PhysRevLett.115.036806,PhysRevLett.115.036807,Yan2017, Chang2017, Chen2017,Kobayashi2017Crossing}. All the proposed nodal line systems are protected by crystallographic symmetry. Some are protected by the coexistence of time-reversal symmetry $T$ and spatial-inversion symmetry $P$, namely $PT$ symmetry \cite{cheng2017body,PhysRevB.93.121113,PhysRevB.93.085427,PhysRevLett.115.036806}.  Some are protected by mirror symmetry or glide symmetry \cite{PhysRevB.90.205136,bzduvsek2016nodal}.

In this work, based on first-principles calculations, we theoretically study a new type of nodal structure, namely nodal net, in two AlB$_2$-type diborides TiB$_2$ and ZrB$_2$, which present a unique combination of properties such as high bond strengths, high melting points, high thermal conductivities, low electrical resistance and low work functions and can be easily synthesized in the laboratory \cite{waskowska2011thermoelastic, okamoto2010anisotropic, kumar2012electronic,wang2011first}. To better understand this complex nodal net structure, we find it including four classes nodal line structures: A, nodal ring in $k_z=0$ plane surrounding $K$ point; B, nodal line in three vertical mirror planes $\sigma_{v1}$, $\sigma_{v2}$ and  $\sigma_{v3}$ (see definition on p. 273 of ~\cite{altmann1994point}); C, nodal line along $\Gamma-A$ starting from a triple point; and D, a single isolated nodal ring at $k_z$=0.5 plane surrounding $A$ point. Furthermore, class-A and class-B nodal lines will cross at a k-point along the $\Gamma-K$ direction.  All three class-B nodal lines in the vertical mirror planes terminate at $A$ point, which is also a termination of the class-C nodal line. So $A$ point is also called a neuxs point~\cite{PhysRevD.59.125015,Heikkila2015Nexus}, which is the termination of several Dirac line nodes. This nodal net structure differs from previously reported 0D and 1D nodal structures which may lead to some new magnetic and electrical transport properties. In addition, the linearly and quadratically dispersed surface Dirac cones are found at the $\bar{K}$ point of the surface BZ for B-terminated and Ti-terminated surfaces of TiB$_2$ respectively.

This paper is organized as follows. In Section.\ref{sec:electronicstructure}, we elucidate the crystal and electronic structure of TiB$_2$. In Section.\ref{sec:nodalnet}, we study the complex nodal net structure both using first-principles calculations and effective $k\cdot p$ model analysis. In Section.\ref{sec:ss}, we study the drumhead surface states both for B-terminated and Ti-terminated surfaces of TiB$_2$, where a linearly and a quadratically dispersed surface Dirac points are found.

\begin{figure}[!htp]
    \centering
    \includegraphics[width=0.5\textwidth]{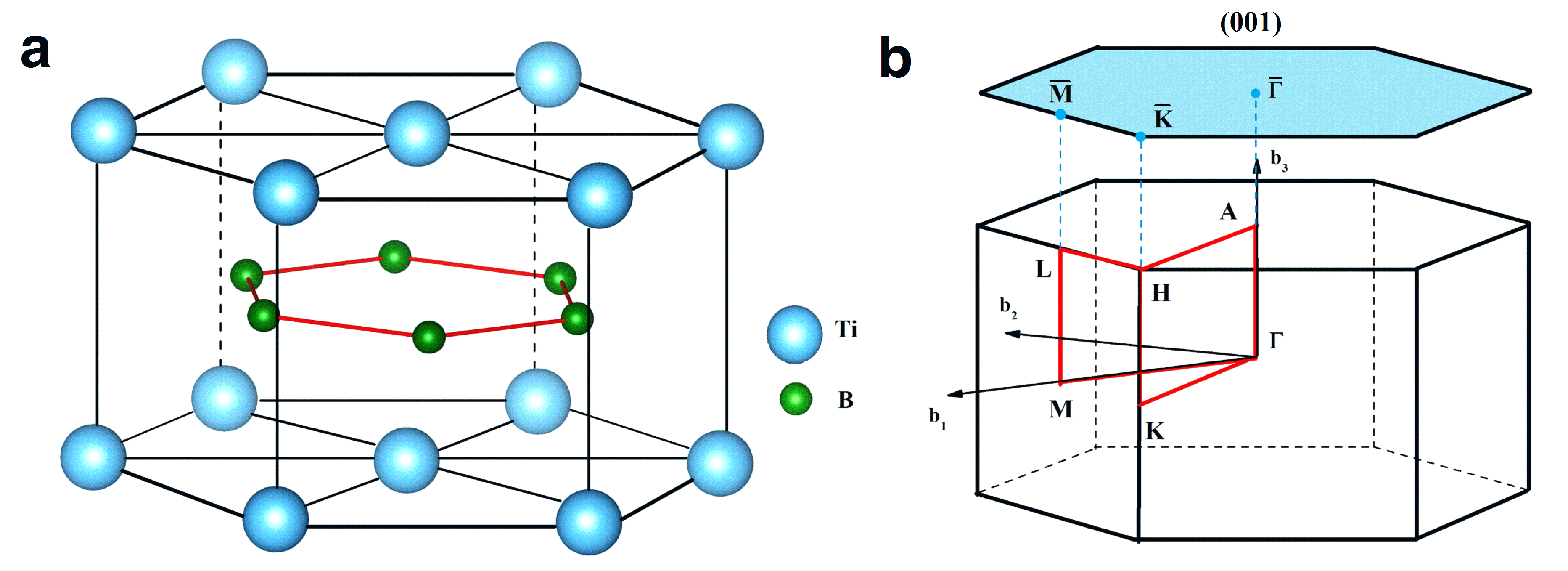}
    \caption{ (color online) Crystal structure and Brillouin Zone (BZ) of AlB$_{2}$-type TiB$_{2}$. (a) Crystal structure of AlB$_{2}$-type TiB$_{2}$ with P6/mmm symmetry. Ti atoms occupy the (0.0, 0.0, 0.0) site, and B atoms occupy the (1/3, 2/3, 1/2) site. The optimized lattice constants are a=b=3.0335 $\r{A}$ and c=3.2263 $\r{A}$. (b) bulk BZ and the projected BZ of the (001) surface. }
    \label{fig1}
\end{figure}

\section{Crystal and band structure of AlB$_{2}$-type TiB$_{2}$}\label{sec:electronicstructure}
The crystallographic data of TiB$_2$ and ZrB$_2$ are obtained from Ref.\cite{kumar2012electronic}. TiB$_2$ and ZrB$_2$ have the same AlB$_2$-type centro-symmetric crystal structures with the space group P6/mmm (191). As verified by calculation, we find TiB$_2$ (Fig.\ref{fig2}) and ZrB$_2$ (Fig \ref{fig5} in Appendix) have similar electronic structures, therefore, we take TiB$_2$ as an example hereafter. As shown in Fig.\ref{fig1}a, it is a layered hexagonal structure with alternating close-packed hexagonal layers of titanium and graphene-like boron layers. The optimized lattice constants are a=b=3.0335 $\r{A}$ and c=3.2263 $\r{A}$, which agree well with the experimental \cite{post1954transition} and other theoretical \cite{milman2001elastic} results.

To study the electronic properties of TiB$_2$, the electronic band structure (BS) is calculated in absence of SOC  (see details in Appendix \ref{append:abinito}), as shown in Fig.\ref{fig2}a. It shows that the valence and conduction bands near the Fermi level exhibit Dirac linear dispersion. There are six band crossing points (also called nodal points (NPs)) located along the H-$\Gamma$, $\Gamma$-A, A-H, K-$\Gamma$, M-K and L-A lines (marked as a to f points in Fig.\ref{fig2}a). It is noticed that these six NPs deviate from the Fermi level about -0.16 eV, 0.5 eV, 0.35 eV, -0.01 eV, -0.28 eV and 0.32 eV, respectively. To investigate the formation mechanism of these crossing points, orbital-character analysis is performed. As shown in Fig.\ref{fig2}a, the Ti 4d states in TiB$_2$ are dominant feature for these six NPs. Specifically, a is dominated by Ti-d$_{xz}$, Ti-d$_{x2-y2}$, and Ti-d$_{z2}$ orbitals; b  is dominated by Ti-d$_{yz}$ and Ti-d$_{z2}$ orbitals; c is dominated by Ti-d$_{xy}$, Ti-d$_{yz}$ and Ti-d$_{x2-y2}$ orbitals; d  is dominated by Ti-d$_{xz}$ and Ti-d$_{z2}$ orbitals; e is dominated by Ti-d$_{xz}$, Ti-d$_{x2-y2}$ and Ti-d$_{z2}$ orbitals, and f is dominated by Ti-d$_{yz}$ and Ti-d$_{x2-y2}$ orbitals, respectively. The Fermi surface (FS) of TiB$_{2}$ is calculated and shown in Fig. \ref{fig2}b and \ref{fig2}c, which shows a lantern-like frame with compensated electron pockets and hole pockets, which is a feature of a topological semimetal and also will lead to non-saturated large positive magnetoresistance~\cite{Ali2014}. The FS calculated in this work agrees well with previous experimental \cite{tanaka1980haas} and theoretical studies \cite{kumar2012electronic}.

In presence of the SOC effect, the crossing points along the H-$\Gamma$, A-H, K-$\Gamma$, M-K and L-A lines are fully gapped, which is common in $PT$ protected systems~\cite{PhysRevLett.115.036806,Fang2016Topological}. SOC makes the crossing points open gaps about 26 meV, 18 meV, 25 meV, 23 meV, and 21 meV [Fig. \ref{fig:gapopen_soc} in Appendix], respectively. The millivolt level gaps indicated that the effect of SOC on the electronic band structure of TiB$_{2}$ is quite weak and can be ignored in experimental work. One thing worth mentioning is that the SOC splitting would generate a Dirac point along the $\Gamma-A$ direction [Fig.\ref{fig:gapopen_soc}b in Appendix] although it happens between the (N+2)'th and (N+3)'th bands, where N is the number of occupied bands at $\Gamma$ point in the BZ .

\begin{figure}[!htp]
    \centering
    \includegraphics[width=0.5\textwidth]{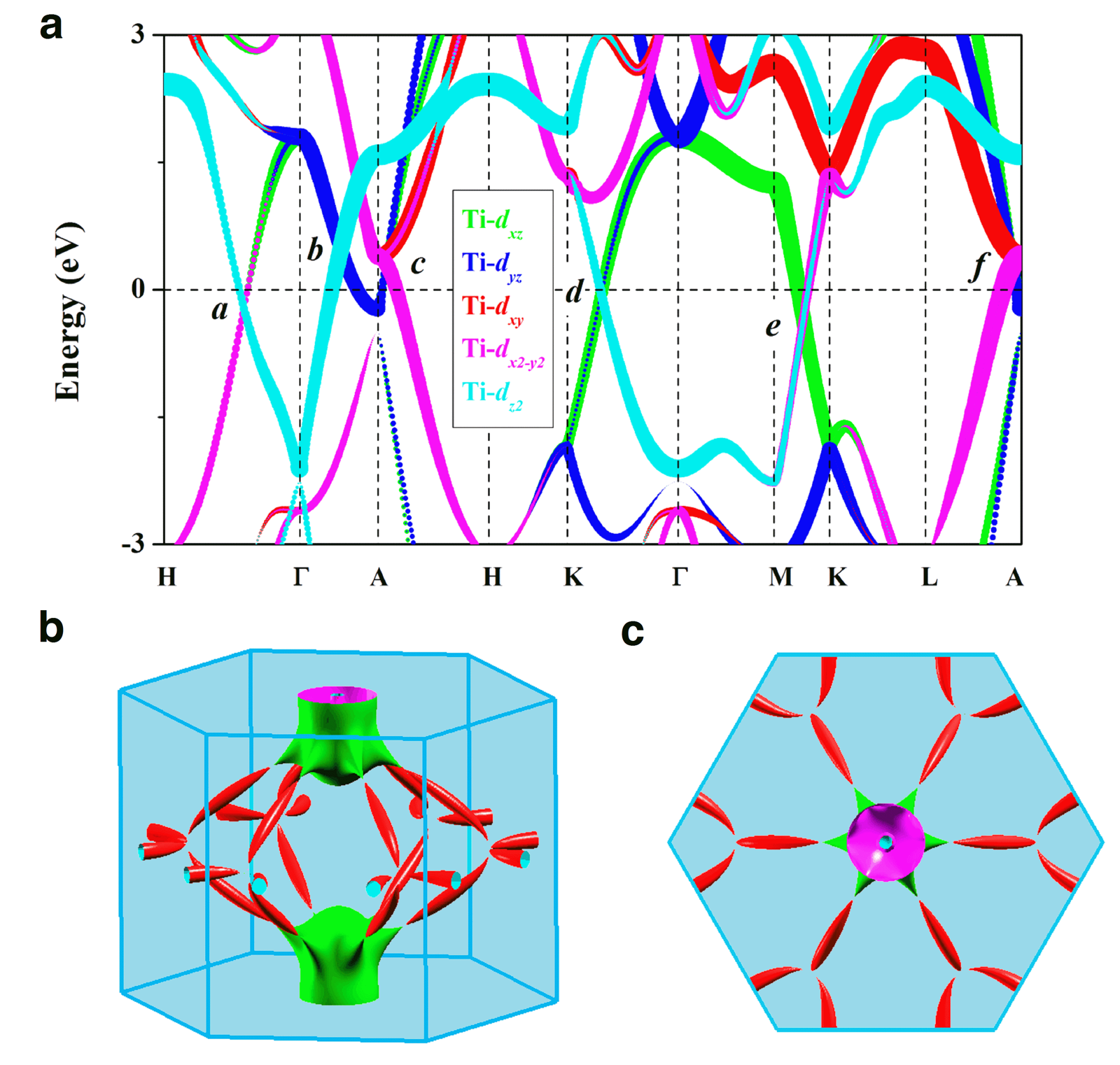}
    \caption{(color online) Electronic energy band and Fermi surface of TiB$_{2}$. (a) Fat-band of TiB$_{2}$. (b) Side view and (c) Top view of the Fermi surface of TiB$_{2}$.}
    \label{fig2}
\end{figure}

\section{Nodal net structure} \label{sec:nodalnet}
\begin{figure}[!htp]
    \centering
    \includegraphics[width=0.46\textwidth]{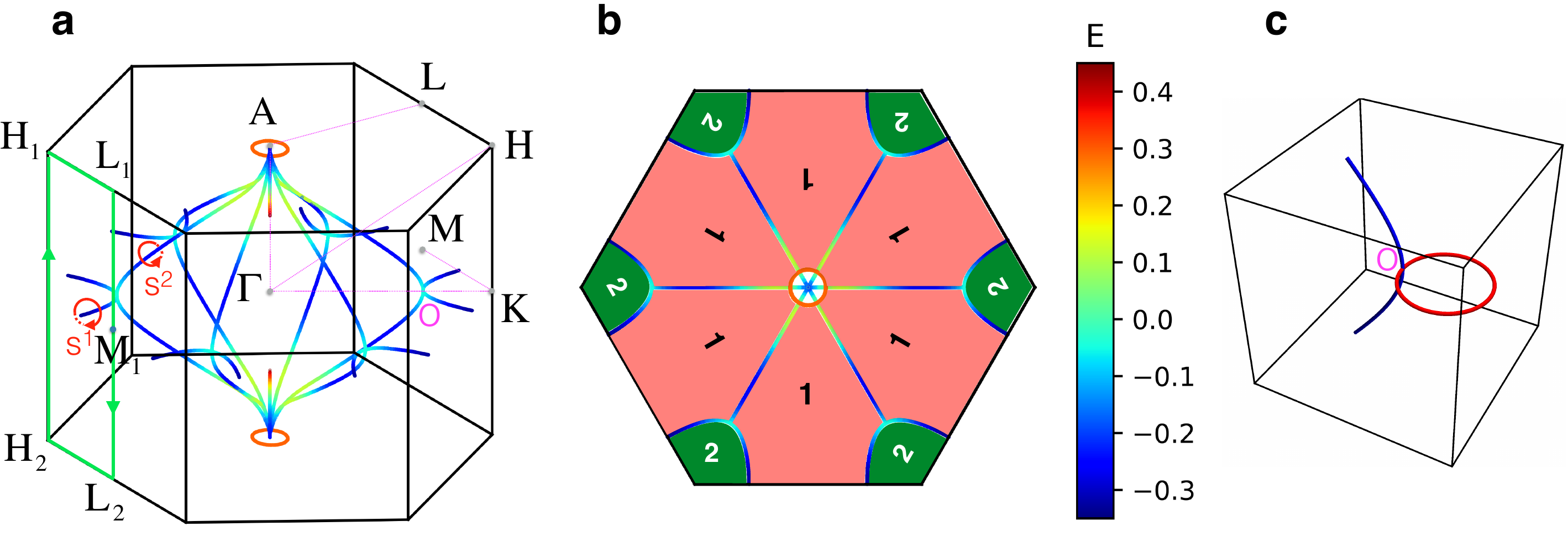}
    \caption{(color online) Nodal-net structure of TiB$_2$. (a), all nodal points in the first BZ. The color indicates the energy of nodal points with reference to the Fermi energy. Two red arrowed circles and the closed green rectangle are used to calculate the Berry phase.  (b), top view of (a), topological number $\nu$ for regions 1 and 2 are 0 and 1, respectively. (c) sketch view of the nodal link of class-A NL (red) and class-B NL (blue), O is the nodal link point.}
    \label{fig3}
\end{figure}

From previous studies, it is known that a nodal line system would have banana-shaped linked FSs~\cite{PhysRevB.80.100508, PhysRevB.94.121108} as the NPs do not usually align at the same energy level. In other words, these banana-shape linked FSs are an indication of the existence of a nodal line structure. There is another clue to find NLSM for a $PT$ symmetry protected system, \emph{i.e.}, if there is a band touching point close to the Fermi level, there should be a nodal line including this point~\cite{Fang2016Topological}. Based on these two clues, and the BS and FS shown in Fig.\ref{fig2}, it is clear that there is a nodal line structure in the TiB$_2$ system. By using the symmetrical Wannier tight binding model~\cite{PhysRevX.6.031003} and a software package WannierTools~\cite{wanniertools}, we found all k-points with zero local energy gap between the N'th and (N+1)'th energy bands, where N is the number of occupied bands at the $\Gamma$ point, \emph{i.e.}: $\Delta({\bf k})= E_{N+1}({\bf k})- E_N({\bf k})=0$. The nodal points are plotted in Fig.\ref{fig3}, from which, the energies of nodal points are not the same which leads to the lantern-like FSs as shown in Fig.\ref{fig2}b.

TiB$_2$ has $PT$ symmetry, which is enough to protect the existence of Dirac nodal lines in the absence of SOC. While, besides $PT$ symmetry, there are another four mirror symmetries $\sigma_h$, $\sigma_{v1}$, $\sigma_{v2}$ and  $\sigma_{v3}$ in the $D_{6h}$ group. Such mirror symmetries will enforce the nodal lines embed on mirror planes. Thus, these nodal lines form a interconnected nodal net structure including four classes nodal lines: A, B, C, and D. In the Appendix, with the aid of DFT calculations, it is verified that class-A,B, and D nodal lines still exist but apart from the previous mirror plane by breaking the  mirror symmetries, however the class-C nodal lines will disappear.

{\it Class-A nodal lines}. Those nodal rings surrounding $K$ point, embed in the $k_z=0$ plane, which is a mirror plane $\sigma_h$ of $D_{6h}$, and is shown as six arcs around $K$ point in Fig.\ref{fig3}. The effective $k\cdot p$ model at $K$ point was constructed within the little group $D_{3h}$, and shown in  Eq. (\ref{eqn:kp-K-twoblock}). When $k_z=0$, $H_{12}=0$, which leads to two uncoupled blocks. Those two eigenvalues of each block lead to one upward and one downward parabola. So, if the energies of the two blocks at $K$ point are different, there will be a nodal ring surrounding $K$ point.

{\it Class-B nodal lines}. Those Weyl nodes sitting on the vertical mirror planes  $\sigma_{v1}$, $\sigma_{v2}$ and  $\sigma_{v3}$, are shown as edges of the lantern in Fig.\ref{fig3}a. We could use the effective $k\cdot p$ model at $\Gamma$ point shown in Eq.(\ref{eqn:kp-gamma}) to prove the existence of a nodal line on such k-planes. For simplicity, we choose $\sigma_{v1} (\sigma_{xz})$, a mirror perpendicular to the y-axis as an example. On this plane, $k_y=0$, then the eigenvalues of Eq.(\ref{eqn:kp-gamma}) are
\begin{align*}
&\epsilon_1=\frac{1}{2}(\alpha+\gamma_1+\gamma_2+\sqrt{(\alpha+\gamma_1-\gamma_2)^2+8\beta^2})\\
&\epsilon_2=\gamma_1-\alpha\\
&\epsilon_3=\frac{1}{2}(\alpha+\gamma_1+\gamma_2-\sqrt{(\alpha+\gamma_1-\gamma_2)^2+8\beta^2})
\end{align*}
where $\alpha=C k_x^2$, $\beta=Dk_x k_z$ and $\gamma_i=E_i+A_i k_x^2+ B_i k_z^2$ with i=1, 2. At $\Gamma$ point, $k_x=0, k_z=0$, $\epsilon_1=E_1$, $\epsilon_2=E_1$, $\epsilon_3=E_2$. It is clear that, along $\Gamma-A$, where $k_x=0$, $\epsilon_1=\epsilon_2$, which is evidence of the existence of class-C nodal lines. The nodal line in classes other than class-C could exist only if $\epsilon_1=\epsilon_3$ or $\epsilon_2=\epsilon_3$. $\epsilon_1=\epsilon_3$ would leads to constraint $E_2- E_1= Ck_x^2$ and $k_x k_z=0$, which results in a touching point $k_z=0, k_y=0, k_x=\sqrt{(E_2-E_1)/C}$. However, the DFT fitting results shows that $E_2-E_1<0$. So there is no touching point between $\epsilon_1$ and $\epsilon_3$. Another possibility, $\epsilon_2=\epsilon_3$, would lead to:
\begin{align}
(C^2+A_2-A_1)k_x^2+(B_2-B_1-D)k_z^2=E_1-E_2 \label{eqn:class-b-constrain}
\end{align}
There are three possibilities arising from the relationship between the parameters in Eq.\ref{eqn:class-b-constrain}
\begin{enumerate}
\item  If $(C^2+A_2-A_1)(E_1-E_2)>0$ and $(B_2-B_1-D)(E_1-E_2)>0$, then there will be an elliptic nodal ring centered at $\Gamma$ point.
\item  If $(C^2+A_2-A_1)(B_2-B_1-D)<0$, then there will be an hyperbolic nodal ring.
\item  If $(C^2+A_2-A_1)(E_1-E_2)<0$ and $(B_2-B_1-D)(E_1-E_2)<0$, there will be no nodal line.
\end{enumerate}
According to the DFT fitting parameters, TiB$_2$ belongs to the first class, which will have a elliptical nodal ring on three vertical mirror planes  $\sigma_{v1}$, $\sigma_{v2}$ and  $\sigma_{v3}$.

Actually, there are another three mirror symmetries $\sigma_{d1}$,  $\sigma_{d2}$ and  $\sigma_{d3}$ in $D_{6h}$. Take $\sigma_{d1}(\sigma_{yz})$ for instance, we could perform the same analysis above by setting $k_x=0$; however, we have to admit that the results are the same as in the $k_y=0$ plane, \emph{i.e.} there will be nodal line in the $k_x=0$ plane. The reason for this is that the $k\cdot p$ model in Eq.(\ref{eqn:kp-gamma}) is up to second order, which will lead to isotopic effects on $k_x$ and $k_y$. Eventually, there are not only  nodal lines in the mirror planes, but also a nodal surface encompassing the $\Gamma$ point. So, in order to degenerate the nodal surface and distinguish between the $k_x=0$ and $k_y=0$ planes, we have to include higher order terms in the $k\cdot p$ model, as discussed in the Appendix.\ref{appendix:kpmodel-K}.

{\it Nodal link of class-A and class-B NLs}. Fig.\ref{fig3}a shows that the class-A NL is linked with class-B NL at O point along $\Gamma-K$ direction, which is also shown in Fig.\ref{fig3}c. $\Gamma-K$ is the overlap of $\sigma_v$ and $\sigma_h$ mirror planes. Eventually, the little group of O is $c_{2v}$ with an additional $PT$ symmetry. The 2-band effective $k\cdot p$ model of O up to 2nd order of ${\bf k}$ is given as
\begin{align}
H(k)= (M+ v k_x +\alpha_1 k_x^2 + \alpha_2 k_y^2+ \alpha_3 k_z^2)\sigma_z + \beta k_yk_z \sigma_x \label{eqn:kp-O}
\end{align}
where $\sigma_x$ and $\sigma_z$ are Pauli-matrices. Since O is a nodal link point, so the mass term $M$ should be zero.  Eigenvalues of Eq.\ref{eqn:kp-O} are
\begin{eqnarray}
E_{\pm}= \pm\sqrt{(v k_x +\alpha_1 k_x^2 + \alpha_2 k_y^2+ \alpha_3 k_z^2)^2+ \beta^2 k_y^2k_z^2}
\end{eqnarray}
The nodal point exists only if $k_y k_z=0$ and $v k_x +\alpha_1 k_x^2 + \alpha_2 k_y^2+ \alpha_3 k_z^2=0$, which leads to two class of NLs, one embeds in $k_z=0$ plane, which belongs to class-A NL, the other one embeds in $k_y=0$ plane, which belongs to class-B NL. It is clearly that both NLs link together at O. Taking class-A as an example, $k_z=0$, and the other condition becomes $\alpha_1 (k_x+\frac{v}{2\alpha_1})^2 + \alpha_2 k_y^2=\frac{v^2}{4\alpha_1}$. It is learned that $v\neq 0$ is the only condition that keeps the NL. $\alpha_1\alpha_2>0$ would lead the NL to be elliptic (class-A NL shown as red ring in Fig.\ref{fig3}c), while $\alpha_1\alpha_2<0$ would lead the NL to be hyperbolic (class-B NL shown as blue line in Fig.\ref{fig3}c).
It is easily to prove that the breaking down of the mirror symmetry would only move the nodal link point O and distort the NLs. However, the breaking down of $PT$ symmetry would gap out all nodal points, because $\sigma_y$ matrix would be introduced to Eq.\ref{eqn:kp-O} then.  So the nodal link of class-A and class-B NLs are robust under the protection of $PT$ symmetry.

{\it Class-C nodal lines}. Actually, such nodal lines are formed by two degenerated bands along the $\Gamma-A$ direction, are protected by $c_{3}$ symmetry, and are shown as a segment between $\Gamma$ and $A$ points in Fig.\ref{fig3}a. Along $\Gamma-A$, there is another topological Fermion, namely the triple point~\cite{PhysRevX.6.031003}, which is a type-A triple point according to the Ref.\cite{PhysRevX.6.031003}'s definition,  because the Berry phase around $\Gamma-A$ is zero. Such a triple point would evolve into a Dirac point in the presence of SOC [Fig.\ref{fig:gapopen_soc}b]. However, the topologically induced SSs of triple points would only happen in a system with SOC, and would not happen in the absence of SOC.

{\it  Class-D nodal lines}. This is an isolated nodal ring at $k_z$=0, and is protected by the $\sigma_h$ mirror symmetry. It is easily deduced from the $k\cdot p$ model (Eq.\ref{eqn:kp-A-block}). The eigenvalues of Eq.\ref{eqn:kp-A-block} are $\epsilon_1^{\pm}=E_1+(A\pm C)(k_x^2+k_y^2)$ and $\epsilon_3^{\pm}=E_3+(B_3\pm G)(k_x^2+k_y^2)$. since $E_1<0$ and $E_3>0$, but $(A\pm C)>0$ and $(B_3+G)*(B_3-G)<0$, $\epsilon_1^{\pm}$ would have a cross-point with $\epsilon_3^{\pm}$ due to its typical band inversion. The resulting nodal line is shown as an orange circle at the top plane of Fig.\ref{fig3}a.

{\it Topological number for the nodal net}. The topological number of DNL is characterized by a quantized $\mathbb{Z}_2$  topological charge $\nu$~\cite{PhysRevLett.115.036806}, which is given by the parity of the Berry phase along a loop $S$ that interlinks with the Dirac ring [red loop in Fig.\ref{fig3}a]. It is verified that $\nu$ is 1 for red loops $S^1$ and $S^2$. The topological charge $\nu$ of red loops $S^1$ is identical to the topological charge of  the green circle in Fig.\ref{fig3}a which is composed of lines H$_2$-H$_1$, H$_1$-L$_1$, L$_1$-L$_2$, and L$_2$-H$_2$. Due to the mirror symmetry, the summation of Berry phases along H$_1$-L$_1$ and L$_2$-H$_2$ is zero. We could define a topological charge $\nu_H$ and $\nu_L$ for  H$_2$-H$_1$ and L$_1$-L$_2$ respectively since they form a closed loop in k-space. From previous studies~\cite{PhysRevLett.115.036806}, The topological number $\nu$ for a time reversal invariant loop which links two parity-invariant momenta $\Gamma_a$ and $\Gamma_b$ is related to the parity of  $\Gamma_a$ and $\Gamma_b$ as
\begin{align}
(-1)^{\nu}= \xi_a\xi_b; \xi_a=\prod_n \xi_n(\Gamma_a)
\end{align}
where $\xi_n(\Gamma_a)$ is the parity for the occupied bands.

So $(-1)^{\nu_L}=\xi_{M_1}\xi_{L_1}$. Since closed loop L$_1$-L$_2$ links $M_1$ and $L_1$ which are the parity-invariant momenta, it is verified by DFT calculations that $\xi_{M_1}=1$ and $\xi_{L_1}=1$ (see details in Appendix.\ref{appendix:toponumber}, \emph{i.e.} $\nu_L=0$. So the topological number for  H$_2$-H$_1$ is $\nu_{H}=(\nu_{S^1}-\nu_L) \ \text{mod}\  2 = 1$. Eventually, the topological numbers are $\nu=0$ and $1$ for regions 1 and 2 shown in Fig.\ref{fig3}b with different colors respectively. There will be odd number of nodal lines between regions 1 and 2 due to the topological number change from 0 to 1.

\begin{figure*}[!htpb]
    \centering
    \includegraphics[width=0.95\textwidth]{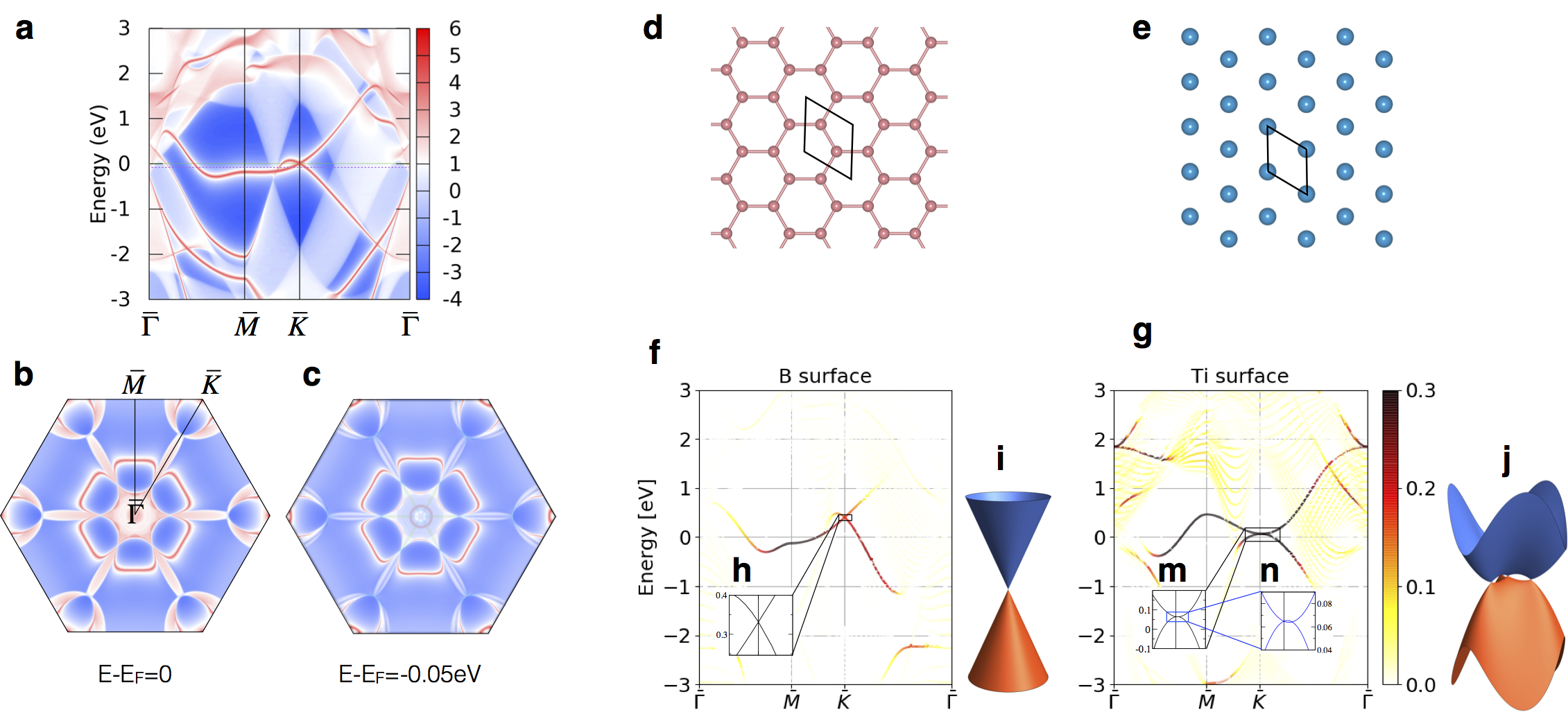}
    \caption{(color online) Surface states analysis of TiB$_{2}$ (001) surface. (a) B-terminated surface state spectrum calculated from the Wannier TB model. The dashed lines are indicative of E-E$_F$=0 and E-E$_F$=-0.05eV. (b) and (c) surface spectrum at fixed energies E-E$_F$=0 and E-E$_F$=-0.05eV which are shown in a subplot. (d) and (e) show surface structure for B-terminated and Ti-terminated surfaces respectively. (g) shows the surface atom weighted band structure which is calculated from VASP for 20 layers of TiB$_{2}$ with Ti(B) sitting in the outer layer. }
    \label{fig4}
\end{figure*}

\section{Drumhead Surface States}\label{sec:ss}
 Based on previous studies \cite{PhysRevLett.115.036806, PhysRevB.93.205132, Kobayashi2017Crossing}, the 1D $\mathbb{Z}_2$ invariant $\nu$ partially guarantees the presence of drumhead surface states. In this section the drumhead surface states on B-terminated and Ti-terminated (001) cleavage surfaces of TiB$_2$ are studied.


{\it B-terminated surface} structure is shown in Fig.\ref{fig4}d, which is a honeycomb lattice like graphene.  By using WannierTools~\cite{wanniertools} and the method of iterative Green's function~\cite{Sancho1985} solution based on a tight-binding model (TBM), the surface state spectrum was calculated, as shown in Fig.\ref{fig4}a-c. In Fig.\ref{fig4}a, it is shown that between $\bar{M}-\bar{K}$ there is a drumhead surface state coming from a "Dirac" point which is the projection of the nodal line. In region 1, those drumhead surface states form a linear Dirac cone at $\bar{K}$ point, which is analogous to the Dirac cone in Graphene. SSs obtained from the TB model usually are used to explain the topological properties. To compare the SS with ARPES experimental data for further use, we use a first-principle calculations for a slab system due to that fact that a real surface system would have charge reconstructions which cannot be described by TBM. So we simulated a 20-layer slab of TiB$_2$ with VASP. The BS is shown in Fig.\ref{fig4}f, in which the color denotes the contributions from the $p_z$ orbital of the surface's boron atoms. Basically, the DFT results are close to the TBM results. The difference is that the zero-energy point of the surface Dirac cone of DFT results is about 0.45 eV higher than that of the TBM results. Since the lattice and the orbital of the boron surface are the same as that of graphene, the effective $k\cdot p$ models at $\bar{K}$ point are the same, and are given as
\begin{align}
H({\bf k})= V_F (k_x \sigma_x + k_y \sigma_y)\label{eqn:lineardirac}
\end{align}
where $V_F$ is the Fermi velocity. By fitted to the DFT calculations, $V_F$ is estimated to $1.5\si{\electronvolt \angstrom}$, which is about $2.28\times 10^5m/s$ in SI units.

{\it Ti-terminated surface} structure is shown in Fig.\ref{fig4}e, which is a hexagonal lattice of the titanium atom. The DFT calculated BS of a 20-layer slab of TiB$_2$ with titanium in the outer surface is shown in Fig.\ref{fig4}g, in which the color denotes contributions from $d_{zx}, d_{yz}$ orbitals of the surface's titanium atoms. There are drumhead SSs coming from the nodal line in this case as same as B-terminated surface; however, the dispersion of SS at $\bar{K}$ is different from the B-terminated surface. In Fig.\ref{fig4}m, it seems that there is a 2D quadratic dispersed Dirac cone at $\bar{K}$ point. However, by zooming into the dispersion at $\bar{K}$ point, it turns out that it is a faked quadratic Dirac cone. Not only there is a linearly dispersed Dirac cone at $\bar{K}$ point, but there is also a linearly dispersed Dirac cone along $\bar{K}-\bar{\Gamma}$. Regarding the point group $C_{3v}$ describing the surface structure, the effective $k\cdot p$ model is obtained as
\begin{align}
H(\bf{k})=\left(\begin{array}{cc}
Bk^2 & Ak_+ +Ck_-^2 \\
Ak_- +Ck_+^2 & Bk^2
\end{array}\right)\label{eqn:sskp-1}
\end{align}
where $k_{\pm}=k_x\pm ik_y$ and $k^2=k_x^2+k_y^2$. When the quadratic terms $B$ and $C$ are missing, then Eq.(\ref{eqn:sskp-1}) reduces to Eq.(\ref{eqn:lineardirac}) which leads to massless linear Dirac dispersion $E(k)=\pm |k|$. In the absence of linear term $A$, Eq.(\ref{eqn:sskp-1}) describes a Dirac point with parabolic energy dispersion. Such a quadratic Dirac point is unstable due to its topological number $\nu$ being trivial. It will split into four linear Dirac points with nontrivial topological number $\nu=1$ under the $C_3$ symmetry ~\cite{Heikkila2015Nexus} which leads to the linear term in Eq.(\ref{eqn:sskp-1}) . Eventually, one is centered at $\bar{K}$ point, the other three  are connected by the $C_3$ symmetry (see Fig.\ref{fig4}j). Fitting to the DFT calculated BS, we obtain the following parameters for a Ti-terminated surface: $A=0.08eV\cdot\AA, B=-0.46 eV\cdot\AA^2$, and $C= 2.3 eV\cdot\AA^2$, in which the linear term is much weaker than the quadratic term.

The surface Dirac cone at $\bar{K}$ of TiB$_2$ with B-terminated and Ti-terminated surfaces are very similar to monolayer and bilayer graphene, which have linearly dispersed and quadratically dispersed Dirac cones respectively. The Berry phase around the linear Dirac cone is $\pi$, while the Berry phase around the quadratically dispersed Dirac cone is $2\pi$. Such differences in the Berry phase would lead to different quantum oscillations.  In Ref.~\cite{PhysRevB.90.161402}, it was proposed that a Dirac point could be observed in monolayer TiB$_2$; however, the Dirac point proposed in this paper could be observed on the surface of a thick slab system rather than a monolayer system.

\section{Conclusion}
In this paper, based on first-principles calculations and model analysis, a novel $PT$ symmetry protected Dirac nodal net state is recognized in AlB$_2$-type TiB$_{2}$ and ZrB$_{2}$ in the absence of SOC. This complex nodal net structure is composed with four classes of NLs: A, B, C, and D, in which, class-A and class-B NLs link together at O along the $\Gamma-K$ direction, three class-B NLs in the vertical mirror planes terminate at A point, which is also a termination of the class-C NL. Several $k\cdot p$ models for these four classes of NLs are constructed under the constraint of their symmetry which confirmed the formation of this nodal net. The topological numbers $\nu$ for different regions in BZ are calculated. It is noted that there are two different dispersed drumhead-like Dirac cones emerging on B-terminated and Ti-terminated surfaces, which are analogous to those of monolayer and bilayer graphene, indicating some novel surface transport properties in TiB$_{2}$ and ZrB$_2$. We believe that this work will guide further progress in understanding the novel properties of TiB$_2$ and ZrB$_2$, and two different terminated surfaces are good platforms to study 2D Dirac fermions. In addition, AlB$_2$-type TiB$_{2}$ and ZrB$_{2}$ can be easily synthesized and provides two prototype materials to study the topological nodal net structure.

{\it Note added.}- Recently, Ref.~\cite{Zhang2017} appeared, discussing some of the topological properties of metal-diboride where the nodal line around $K$ is the Class-A nodal line in the nodal net of this work.

\section{Acknowledge}
We acknowledged helpful discussions with X.Dai.  X.F and B.W were supported by the National Natural Science Foundation of China (NSFC-51372215), C.Y and Z.S were supported by National Natural Science Foundation of China, the National 973 program of China (Grant No. 2013CB921700), Q.W was supported by Microsoft Research, and the Swiss National Science Foundation through the National Competence Centers in Research MARVEL and QSIT.

\appendix
\section{Computational methods}\label{append:abinito}
In this work, the electronic properties for AlB$_2$-type TiB$_2$ are studied by using density functional theory (DFT) \cite{PhysRev.136.B864,PhysRev.140.A1133} as implemented in the Vienna Ab initio Simulation Package (VASP) \cite{kresse1996efficiency,PhysRevB.54.11169,PhysRevB.59.1758}. The exchange correlation functional of Perdew-Burke-Emzerhof generalized gradient approximation (GGA-PBE) \cite{PhysRevB.50.17953,PhysRevLett.77.3865} are performed. The standard version of PBE pseudo-potential is adopted in this work explicitly treating four valence electrons for the Ti atom (4d$^3$5s$^1$) and three valence electrons for the B atom (3s$^2$3p$^1$). A cutoff energy of 500 eV and an 11$\times$11$\times$9 k-mesh  are used to perform the bulk calculation. The conjugate-gradient algorithm is used to relax the ions, and the convergence thresholds for total energy and ionic force component are chosen as 1$\times$10$^{-7}$ eV and 0.001 eV/$\r{A}$.

For the slab calculations (Fig.\ref{fig4}f-g), the thickness of the B-terminated slab is 20 layers of titanium and 21 layers of boron, while the thickness of the Ti-terminated slab is 20 layers of titanium and 19 layers of toron. A $26\times 26\times 1$ $\Gamma$ centered k mesh and a 14$\AA$-thick vacuum are used in the DFT simulations. The surface are fully relaxed with energy convergence up to 1$\times$10$^{-7}$ eV and force up to 0.001 eV/$\r{A}$.

The nodal-net searching and surface states spectrum calculations shown in Fig.\ref{fig4}a-c are done using the open-source software WannierTools~\cite{wanniertools} which is based on Wannier tight binding model (WTBM) constructed with Wannier90~\cite{mostofi2008wannier90}. Ti $s, p, d$, and B $s, p$ orbitals are used as initial projectors for WTBM construction. WTBMs constructed with Wannier90 do not exactly fulfil all crystal symmetries which is very important for nodal points searching because usually nodal points are protected by crystal symmetries except Weyl points. The WTBM is symmetrized to be compatible with the crystal symmetry using the method described in Red.~\cite{PhysRevX.6.031003}.

\begin{figure*}[!htp]
    \centering
    \includegraphics[width=\textwidth]{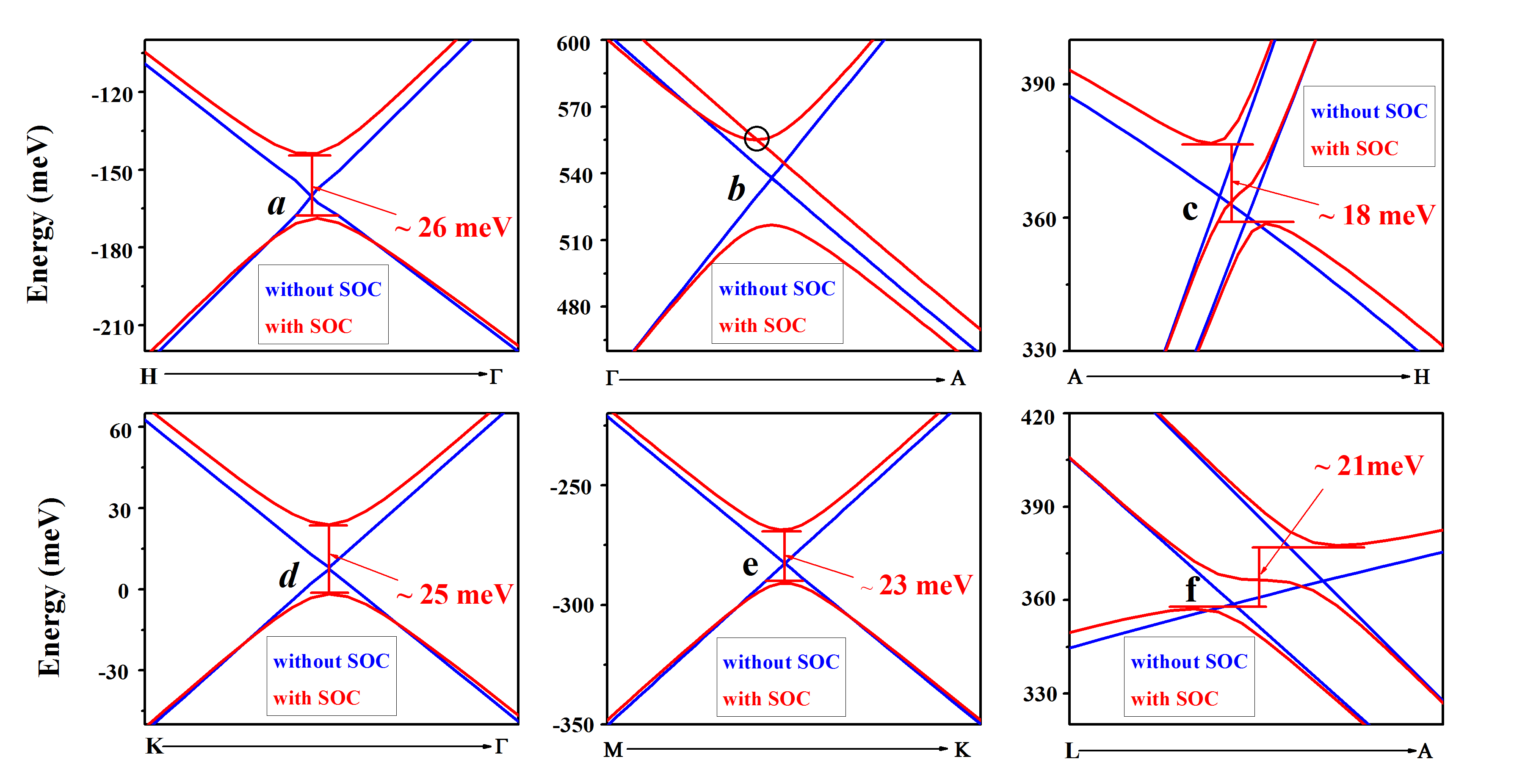}
    \caption{(color online) Energy bands comparison between without SOC and with SOC close to the six nodal points. The triple point along $\Gamma-A$ in the absence of SOC degenerates to a Dirac point between $(N+2)$'th and $(N+3)$'th band which is marked by black circle in panel b when SOC is included. \label{fig:gapopen_soc}}
\end{figure*}

\section{band structure and Fermi surface of AlB$_{2}$-type ZrB$_{2}$}
\begin{figure}[!htp]
    \centering
    \includegraphics[width=0.5\textwidth]{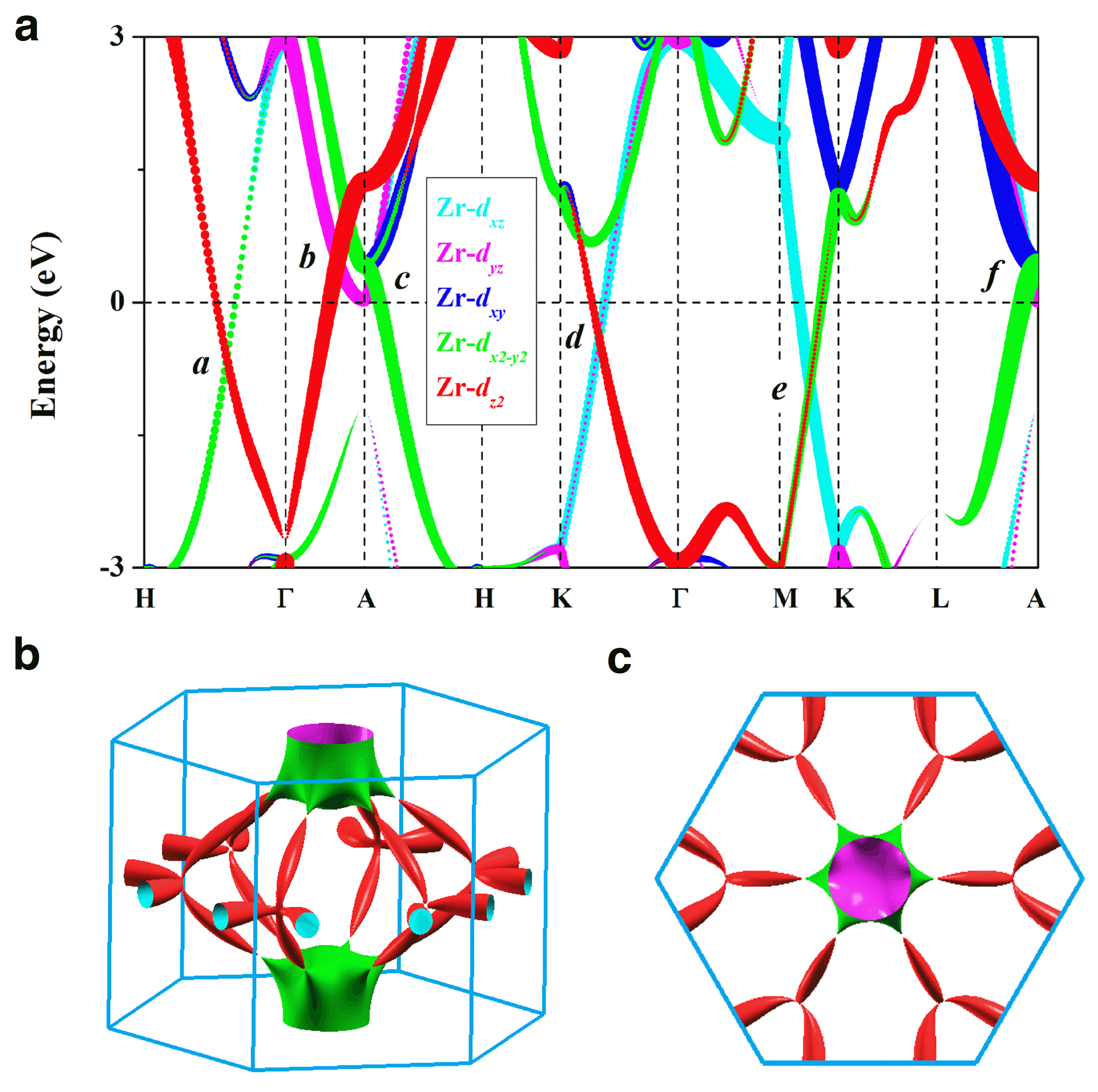}
    \caption{(color online) Electronic energy band and Fermi surface of AlB$_{2}$-type ZrB$_{2}$. (a) Fat-band of AlB$_{2}$-type ZrB$_{2}$. (b) Side view and (c) Top view of the Fermi surface of AlB$_{2}$-type ZrB$_{2}$. }
    \label{fig5}
\end{figure}

In general, ZrB$_{2}$ is often compared with TiB$_{2}$. The structure of AlB$_2$-type ZrB$_2$ also has hexagonal structure with a space group of P6/mmm (No.191). Its optimized lattice constants are a=b=3.1748(3) $\r{A}$ and c=3.5579(7) $\r{A}$, which is slightly larger than those of TiB$_{2}$. As shown in Fig.\ref{fig5}a, there are also six band crossing points along the high-symmetry path similar with TiB$_{2}$. Among them, Zr 4d states are the main contribution orbitals for these six band crossing points. The Zr-d$_{xz}$ and Zr-d$_{xy}$ orbitals are much higher than the Fermi level. From Figs. \ref{fig5}b and c, the Fermi surface of ZrB$_2$ is sightly different from  that of TiB$_2$. The circling surface around A point is disappeared in the Fermi surface of  ZrB$_2$. The main part of the lantern-like Fermi surface of ZrB$_2$ is basically consistent with the Fermi surface of TiB$_2$, which also shows a nodal-net feature.

\section{Nodal net structure of TiB$_2$ without vertical mirror planes}
\begin{figure}[!htp]
    \centering
    \includegraphics[width=0.45\textwidth]{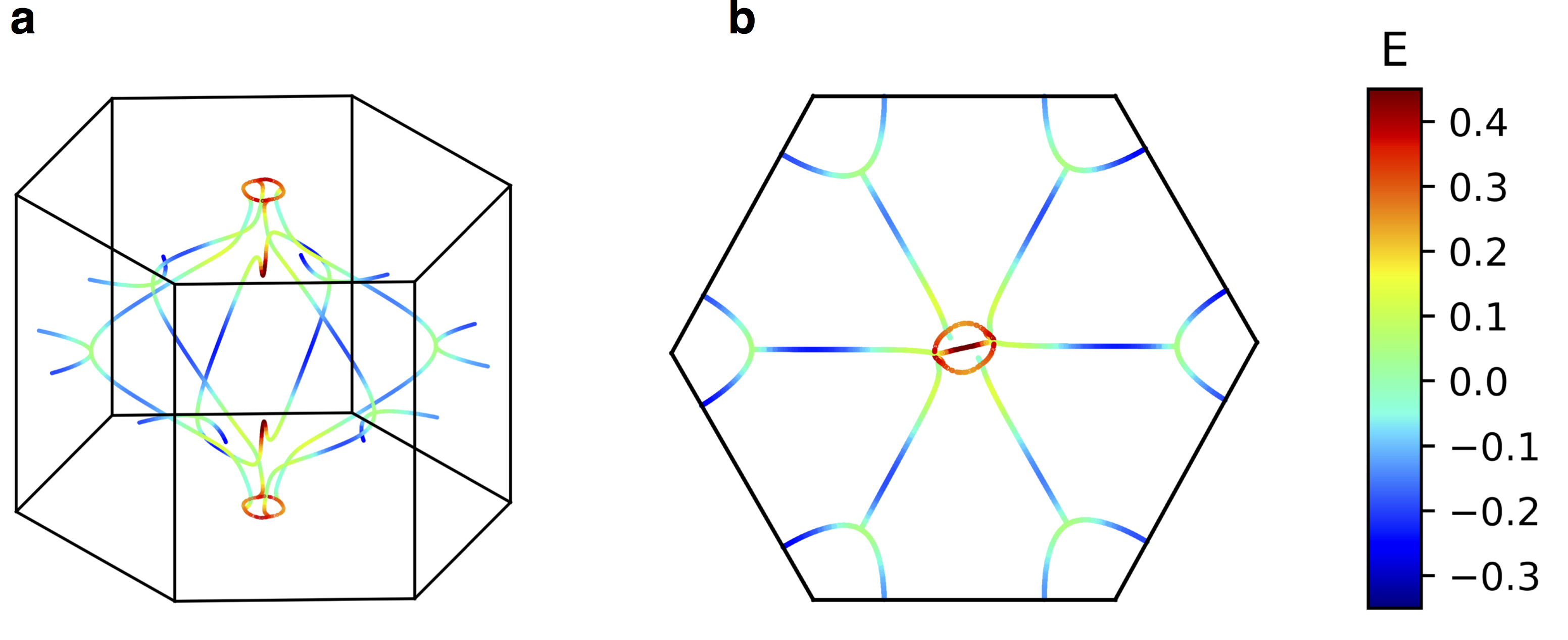}
    \caption{(color online) Nodal net structure of TiB$_2$ without vertical mirror planes}
    \label{fig:nodalnetstrain}
\end{figure}

Since the nodal net of TiB$_2$ is $PT$ symmetry protected, any perturbations that preserve the inversion symmetry can only distort the nodal-net but not destroy them. As an example,we apply a uniaxial strain (compression 1$\%$) along the [100] crystal direction. After deformation, the structure of AlB$_2$-type TiB$_2$ belongs to space group of C2/m (No. 12), which just preserves the M$_z$ mirror reflection symmetry. As a result, as shown in Fig.\ref{fig:nodalnetstrain}, it is found that the class-A nodal line still embeds in the $k_z=0$ plane, one of the class-B nodal line become isolated with the other two class-B nodal lines which link with the class-D nodal line that embeds in the $k_z=0.5$ plane, while the class-D nodal line along $\Gamma-A$ disappears because it is protected by $C_3$ symmetry which is destroyed under such strain. Thus, the nodal net in TiB$_2$ is robustly stable, which does not requires the protection of mirror symmetry.

\section{Effective $k\cdot p$ model}
We derive several $k\cdot p$ models describing the bulk bands in the vicinity of $\Gamma$, A, and K in the 3D BZ, and a $k\cdot p$ for surface states at $\bar{K}$ point in 2D BZ. The $k\cdot p$ models are used to get a better understanding of the surface states and the nodal net structures, and would be useful for further investigations of Landau level and quantum transport properties.

The $k\cdot p$ models were calculated using the kdotp\_symmetry code, which implements the method described in Ref.~\cite{Gresch2017}, basically, there are two things that should be prepared before applying kdotp\_symmetry.  Firstly, we identify the little group $G$ of a high symmetry point $K_0$ of which we want to construct a low energy effective model, and get the generators $R$ of little group $G$. Secondly, we identify the representations of $R$ on the basis of the eigenvectors at the selected high-symmetry points. Then kdotp\_symmetry will produce the $k\cdot p$ model under the constraint
\begin{align}
D(R)H(k)D^{\dag}(R)= H(R(k))\label{eqn:symmetry}
\end{align}
where D(R) is the representation matrix of symmetry operator $R$.

\subsection{$k\cdot p$ model at $\Gamma$ point}
The little point group at Gamma point of bulk TiB$_2$ is $D_{6h}$ plus TR symmetry (see Table 76 of Ref.\cite{Koster1962}). There are three generators of $D_{6h}$ including spatial-inversion $I (-x, -y, -z)$, 2-fold rotation $c_{2y}(-x, y, -z)$ and 6-fold rotation $c_{6z}(\frac{x}{2}+\frac{\sqrt{3}}{2}y, -\frac{\sqrt{3}}{2}x+\frac{y}{2}, z)$. From the fat-band analysis of Fig.\ref{fig2}, the relevant bands come from $d_{z^2}$ which belongs to the $A_{1g}$ representation, $d_{zx}$ and $d_{yz}$ orbitals which form the basis of its $E_{1g}$ representation. The representations of group generators according to the symmetrical basis \{($d_{zx}$+i$d_{yz}$)/$\sqrt2$, ($d_{zx}$-i$d_{yz}$)/$\sqrt2$ , $d_{z^2}$\} are given by
\begin{align}
&D(I)= \text{diag}\{ 1, 1, 1 \}\nonumber\\
&D(c_{6z})= \text{diag}\{-e^{i2\pi/3}, -e^{-i2\pi/3}, 1\} \\
&D(c_{2y})=\left(\begin{array}{ccc}
0 & 1 & 0  \\
1 & 0 & 0 \\
0 & 0 & 1
\end{array}\right)\\
&D(\text{TR})=\left(\begin{array}{ccc}
0 & 1 & 0  \\
1 & 0 & 0 \\
0 & 0 & 1
\end{array}\right)K
\end{align}
where, $K$ is the complex conjugation operator. Considering these symmetries, and the constraint of Eq.(\ref{eqn:symmetry}), a 3-band model up to second order of k around $\Gamma$ point for bulk TiB$_2$ is given by
\begin{align}
&H(\bf{k})= \left(\begin{array}{ccc}
\varepsilon_1({\bf k}) & Ck_-^2 & Dk_-k_z \\
Ck_+^2& \varepsilon_1({\bf k}) & Dk_+k_z \\
Dk_+ k_z & Dk_-k_z & \varepsilon_2({\bf k})
\end{array}\right)\label{eqn:kp-gamma}
\end{align}
where $\varepsilon_1({\bf k})=E_1+A_1(k_x^2+k_y^2) +B_1 k_z^2$, $\varepsilon_2({\bf k})= E_2+A_2(k_x^2+k_y^2)+ B_2k_z^2$.

As mentioned in the main text, Eq.(\ref{eqn:kp-gamma}) with only second order momentum $k$ would lead to a nodal surface other than the nodal line structure. To distinguish the difference between $\Gamma-K$ and $\Gamma-M$ directions, we have to introduce the sixth order of $k_x$, $k_y$ in the $k_z=0$ plane, and introduce a fourth order of $k_x$, $k_y$ in the off-diagonal part, so eventually, the new $k\cdot p$ model is given by
\begin{align}
&H(\bf{k})= \left(\begin{array}{ccc}
\varepsilon_1({\bf k}) & Ck_-^2+ F k_+^4 & Dk_-k_z \\
Ck_+^2+ F k_-^4& \varepsilon_1({\bf k}) & Dk_+k_z \\
Dk_+ k_z & Dk_-k_z & \varepsilon_2({\bf k})
\end{array}\right)\label{eqn:kp-gamma-sixth}
\end{align}
where $\varepsilon_1({\bf k})=E_1+A_1(k_x^2+k_y^2) +B_1 k_z^2$, $\varepsilon_2({\bf k})= E_2+A_2(k_x^2+k_y^2)+ B_2k_z^2+ L (k_x^2+k_y^2)^2+ M (k_+^6+k_-^6)$.

\begin{figure}[!htp]
    \centering
    \includegraphics[width=0.45\textwidth]{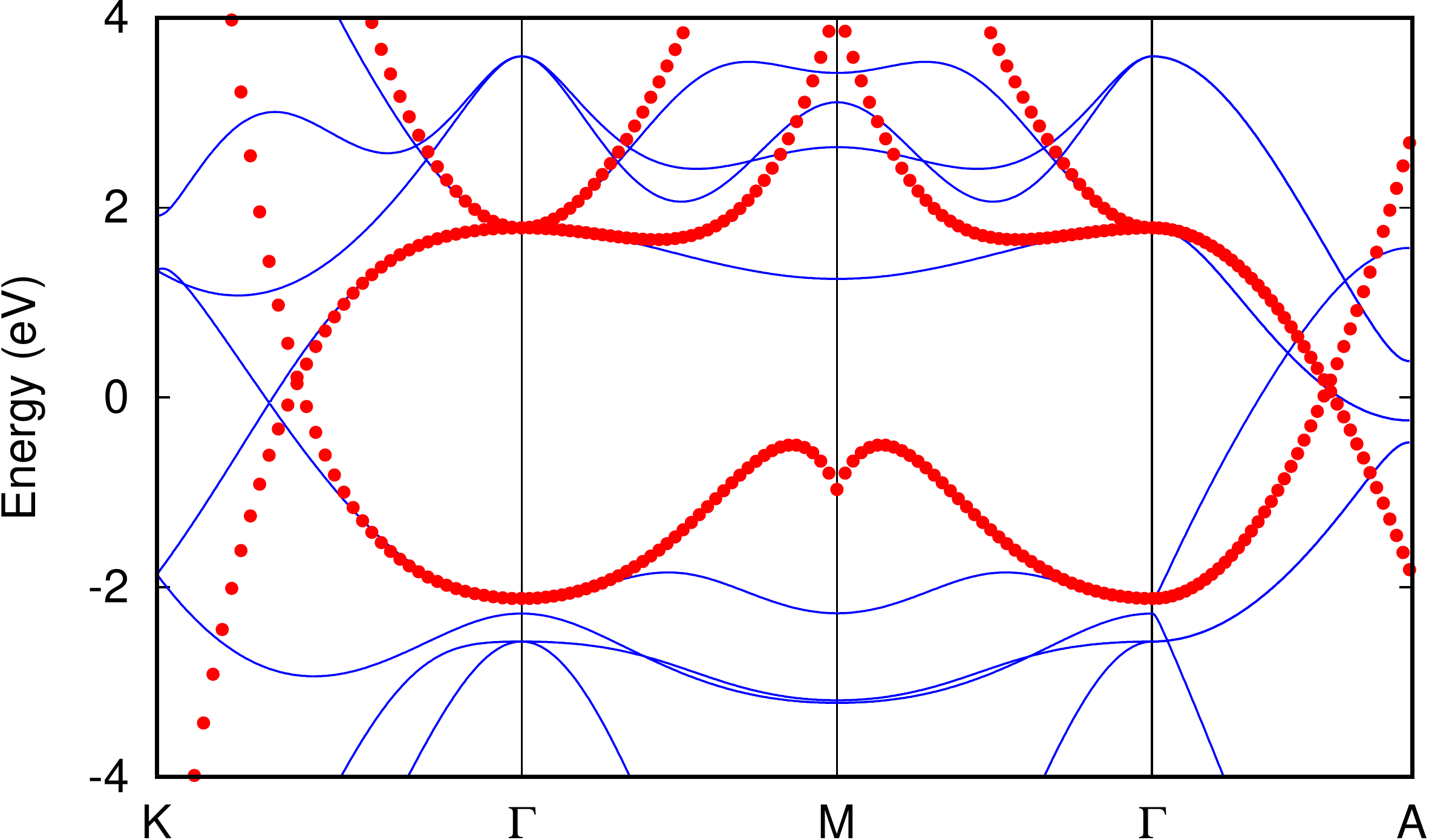}
    \caption{(color online) Comparison between DFT bands and the bands from the $k\cdot p$ model [Eq.(\ref{eqn:kp-gamma-sixth})] at $\Gamma$ point of TiB$_2$. Blue lines come from DFT calculation, and red dotted lines come from kp model. }
    \label{fig:kp-dft-gamma}
\end{figure}

By fitting to the DFT band structure of TiB$_2$, the parameters in Eq.(\ref{eqn:kp-gamma-sixth}) are obtained:  $E_1= 1.787 \si{\electronvolt}$, $B_1=-3.8 \si{\electronvolt \angstrom}^2$, $A_1=2.6 \si{\electronvolt \angstrom}^2$, $E_2=-2.12 \si{\electronvolt}$, $A_2=1.63 \si{\electronvolt \angstrom}^2$, $B_2=5.1 \si{\electronvolt \angstrom}^2$, $L=1.3 \si{\electronvolt \angstrom}^4$,  $C=3.55 \si{\electronvolt \angstrom}^2$, $M=0.65 \si{\electronvolt \angstrom} ^6$, $F=1.83 \si{\electronvolt \angstrom}^4$, and $D=5.1 \si{\electronvolt \angstrom}^2$. The comparison between DFT bands and $k\cdot p$ bands is shown in Fig.\ref{fig:kp-dft-gamma}

\begin{figure}[!htp]
    \centering
    \includegraphics[width=0.45\textwidth]{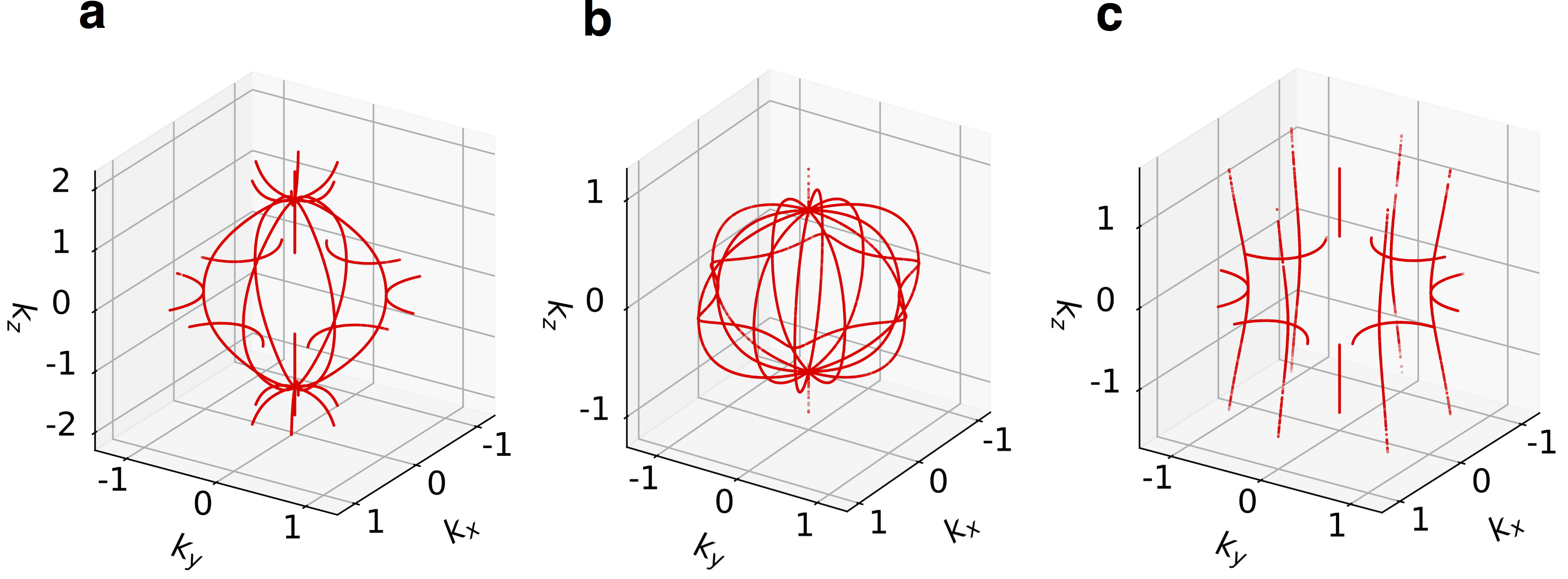}
    \caption{(color online) Nodal line structure obtained from Eq.(\ref{eqn:kp-gamma-sixth}) at $\Gamma$ point of TiB$_2$ with different $M, F$ and $D$ parameters. (a),  $M=0.65 \si{\electronvolt \angstrom}^6$, $F=1.83 \si{\electronvolt \angstrom}^4$, $D=5.1 \si{\electronvolt \angstrom}^2$. (b),  $M=0$, $F=1.18 \si{\electronvolt \angstrom}^4$, $D=2.53 \si{\electronvolt \angstrom}^2$. (c),  $M=0.65 \si{\electronvolt \angstrom}^6$, $F=7.6 \si{\electronvolt \angstrom}^4$, $D=5.1 \si{\electronvolt \angstrom}^2$.  }
    \label{fig:nodalline-kp-gamma}
\end{figure}

The nodal line structure around $\Gamma$ point calculated from Eq.\ref{eqn:kp-gamma-sixth} with the DFT fitted parameters is shown in Fig.\ref{fig:nodalline-kp-gamma}a. It is shown that the nodal net close to $\Gamma$ point is very similar to Fig.\ref{fig3}, the class-A, class-B, and class-C nodal lines are captured successfully; however the nodal line beyond the nexus point $A$ is not captured. That is because the nexus point $A$ is at the boundary of the BZ, which is related to an infinity in the $k\cdot p$ model. The position of the nexus point $A$ could be tuned by changing $F$ in Eq.\ref{eqn:kp-gamma-sixth}. In Fig.\ref{fig:nodalline-kp-gamma}c, it is shown that the nexus point would disappear if $F$ was very large.  The nodal line in the $\sigma_d$ mirror plane could be shown if the fourth and sixth order terms in Eq.\ref{eqn:kp-gamma-sixth} become smaller (Fig.\ref{fig:nodalline-kp-gamma}b).

\subsection{$k\cdot p$ model at K point}\label{appendix:kpmodel-K}
The little group at K point in the TiB$_2$ is $D_{3h}$ (see Table 65 of Ref.\cite{Koster1962}), of which there are three generators including horizontal mirror $\sigma_h (x, y, -z)$, 2-fold rotation $c_{2x} (x, -y, -z)$ and 3-fold rotation $c_{3z} (-\frac{x}{2}-\frac{\sqrt{3}}{2}y, \frac{\sqrt{3}}{2}x-\frac{y}{2}, z)$. Around K point, the relevant representations are $\Gamma_5, and \Gamma_6$,  of which the basis are $\{ d_{zx}, d_{yz}\}$ and  $\{ d_{xy}, d_{x^2-y^2}\}$. Taking the symmetrical orbitals \{ $|Y_2^1\rangle$, -$|Y_2^{-1}\rangle$\} and \{ $-|Y_2^{2}\rangle , |Y_2^{-2}\rangle$\} as a basis, where $|Y_l^m\rangle$ is the complex Spherical harmonic function, the representations of the generators are given by
\begin{align}
&D(\sigma_h)= \text{diag}\{-1, -1\}\oplus  \text{diag}\{1, 1\}\\
&D(c_{2x})= \left(\begin{array}{cc}
0 & -1 \\
-1 & 0
\end{array}\right)\oplus
\left(\begin{array}{cc}
0 & -1 \\
-1 & 0
\end{array}\right)\\
&D(c_{3z})= \left(\begin{array}{cc}
e^{-2i\pi/3} & 0 \\
0 & e^{2i\pi/3}
\end{array}\right)\oplus  \left(\begin{array}{cc}
e^{2i\pi/3} & 0 \\
0 & e^{-2i\pi/3}
\end{array}\right)
\end{align}
Considering these symmetries above and the constraint of Eq.(\ref{eqn:symmetry}), a 4-band model up to second order of k around ${K}$ point in bulk TiB$_2$ is given by
\begin{small}
\begin{align}
&H(k)= \nonumber\\
&\left(
\begin{array}{cccc}
\varepsilon(k)  & C k_+ +D k_-^2  & i E k_z k_+  & F k_z \\
C k_-+ D k_+^2  & \varepsilon(k)  & F k_z  & -i E k_z k_- \\
-i E k_z k_-  & F k_z & \varepsilon'(k) & C' k_-+ D' k_+^2 \\
F k_z & i E k_z k_+  & C' k_+ + D' k_-^2 & \varepsilon'(k)
\end{array}
\right)\label{eqn:kp-K}
\end{align}
\end{small}
where $\varepsilon(k)= E_0+ A(k_x^2+ k_y^2)+ Bk_z^2$ and  $\varepsilon'(k)= E_0'+ A'(k_x^2+ k_y^2)+ B'k_z^2$. The fitted parameters of TiB$_2$ are  $E_0 = -1.8651\si{\electronvolt}$, $E_0'= 1.3387\si{\electronvolt}$,
$A = 2.55\si{\electronvolt \angstrom}^2$, $A' =-16.5\si{\electronvolt \angstrom}^2$,  $C = 3.7\si{\electronvolt \angstrom}$, $C' =-1.66\si{\electronvolt \angstrom}$, $D =-0.58\si{\electronvolt \angstrom}^2$, $D' = 18.8\si{\electronvolt \angstrom}^2$, $B = 8.73\si{\electronvolt \angstrom}^2$, $B' = -6.37\si{\electronvolt \angstrom}^2$, $E = 24.9\si{\electronvolt \angstrom}^2$, $F = 6.84\si{\electronvolt \angstrom}$. The comparison between DFT bands and $k\cdot p$ bands is shown in Fig. \ref{fig:kp-dft-K}.

In order to analyse the nodal line structure, Eq.(\ref{eqn:kp-K}) can be written into two blocks
\begin{align}
H(k)= \left(
\begin{array}{cccc}
H_{11} & H_{12} \\
H_{12}^{\dag} & H_{22}
\end{array}
\right)\label{eqn:kp-K-twoblock}
\end{align}
where
\begin{align}
&H_{11}=
&\left(
\begin{array}{cccc}
\varepsilon(k)  & C k_+ +D k_-^2  \\
C k_-+ D k_+^2  & \varepsilon(k)\end{array}
\right) \\
&H_{22}=
&\left(
\begin{array}{cccc}
\varepsilon'(k)  & C' k_+ +D' k_-^2  \\
C' k_-+ D' k_+^2  & \varepsilon'(k)\end{array}
\right) \\
&H_{12}=
&\left(
\begin{array}{cccc}
i Ek_z k_+ & F k_z  \\
Fk_z  & -i Ek_z k_-
\end{array}
\right)\
\end{align}

\begin{figure}[!htp]
    \centering
    \includegraphics[width=0.45\textwidth]{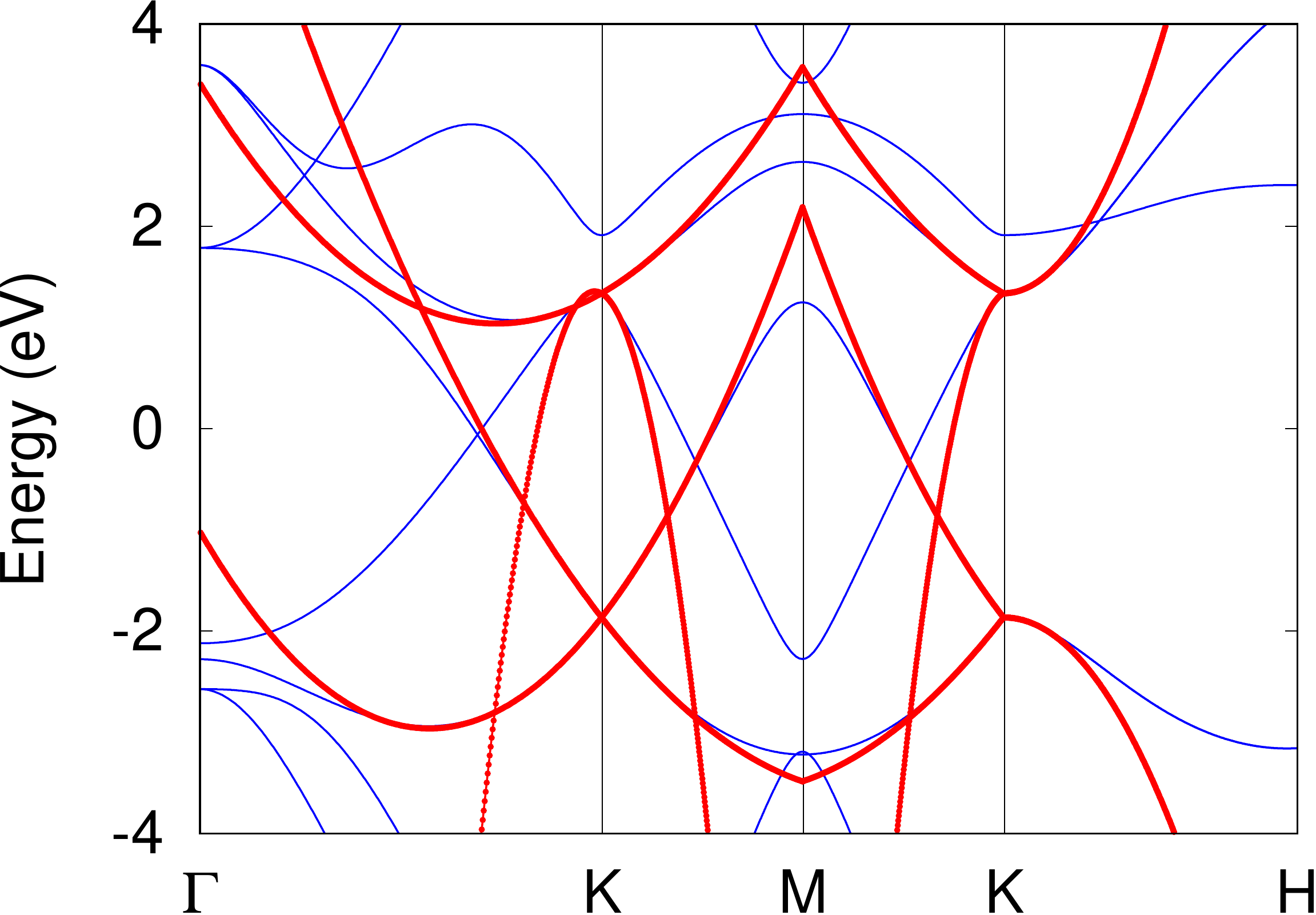}
    \caption{(color online) Comparison between DFT bands and the bands from the $k\cdot p$ model [Eq.(\ref{eqn:kp-K})] at $K$ point of TiB$_2$. Blue lines come from DFT calculation, and red dotted lines come from $k\cdot p$ model. }
    \label{fig:kp-dft-K}
\end{figure}

With the fitted parameters, we find that our $k\cdot p$ model not only describes the nodal line surrounding $K$ point, but can also predict part of the nodal line in the vertical mirror plane.
\begin{figure}[!htp]
    \centering
    \includegraphics[width=0.45\textwidth]{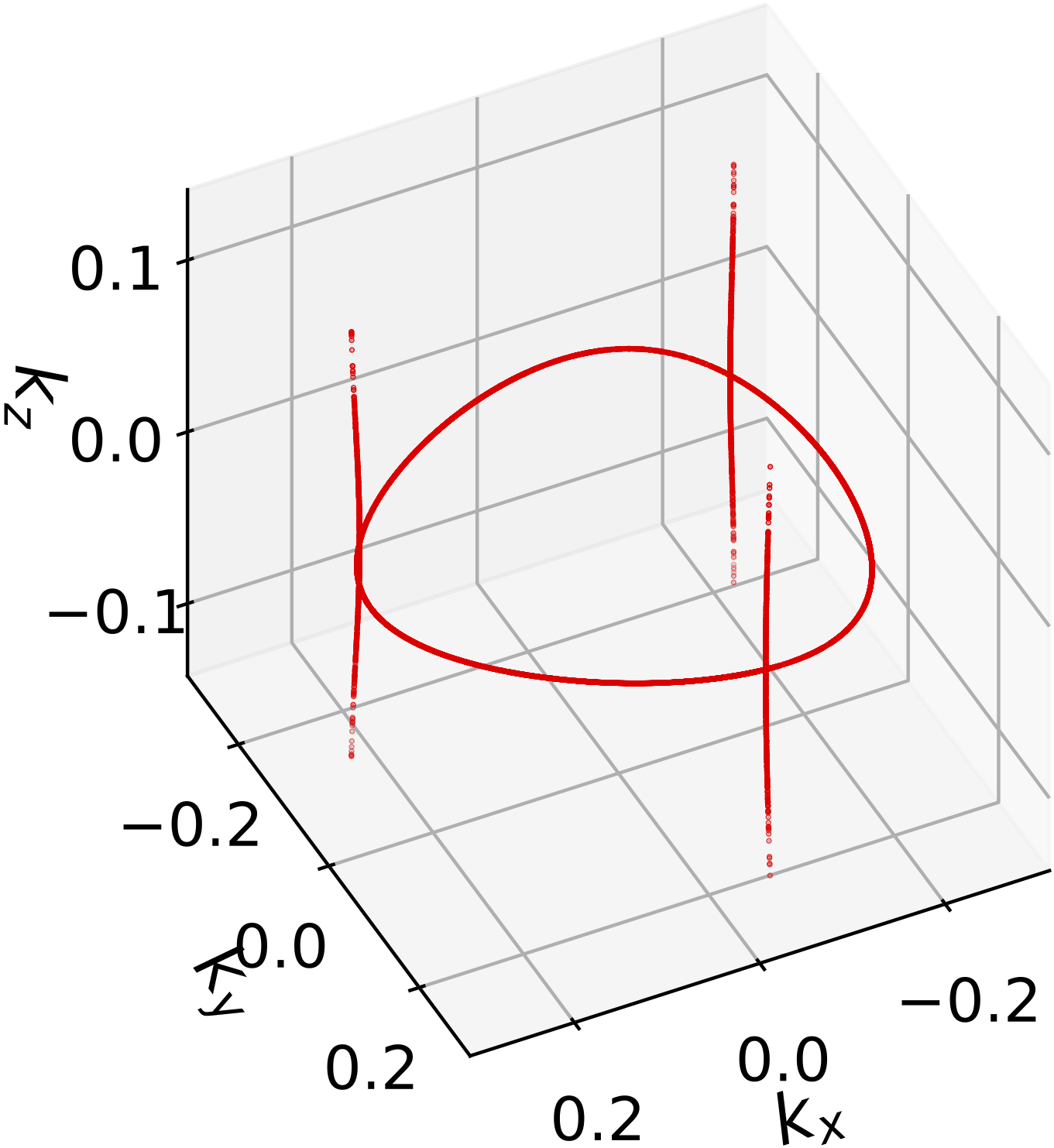}
    \caption{(color online) Nodal line structures close to $K$ point calculated from $k\cdot p$ model (Eq.\ref{eqn:kp-K}). The coordinates are relatively to $K$ point. }
    \label{fig:Nodesline-K}
\end{figure}

\subsection{$k\cdot p$ model at A point}
For the little group at A $D_{6h}$ (see Table 76 of Ref.\cite{Koster1962}),
relevant representations are  $A_{1g}$ $E_{1g}$ and $E_{2g}$, which constitutes \{$d_{z^2}$\}, \{$d_{zx}$, $d_{yz}$\} and\{$d_{xy}$, $d_{x^2-y^2}$\}. By symmetrization of those orbitals according to the $D_{6h}$ group, the symmetrical basis are chosen as $|Y_2^0\rangle$, $|Y_2^1\rangle$, -$|Y_2^{-1}\rangle$, -$|Y_2^2\rangle$ and $|Y_2^{-2}\rangle$, and the related representations of its generators are given by
\begin{align}
&D(I)= \text{diag}\{1, 1, 1, 1, 1\}\\
&D(c_{2y})=1\oplus
\left(\begin{array}{cc}
0 & 1 \\
1 & 0
\end{array}\right)\oplus
\left(\begin{array}{cc}
0 & -1 \\
-1 & 0
\end{array}\right)\\
&D(c_{6z})=\text{diag}\{1, -e^{2i\pi/3}, -e^{-2i\pi/3}, e^{-2i\pi/3}, e^{2i\pi/3}\}\\
&D(\text{TR})=1\oplus
\left(\begin{array}{cc}
0 & 1 \\
1 & 0
\end{array}\right)\oplus
\left(\begin{array}{cc}
0 & -1 \\
-1 & 0
\end{array}\right)K
\end{align}

\begin{figure}[!htp]
    \centering
    \includegraphics[width=0.45\textwidth]{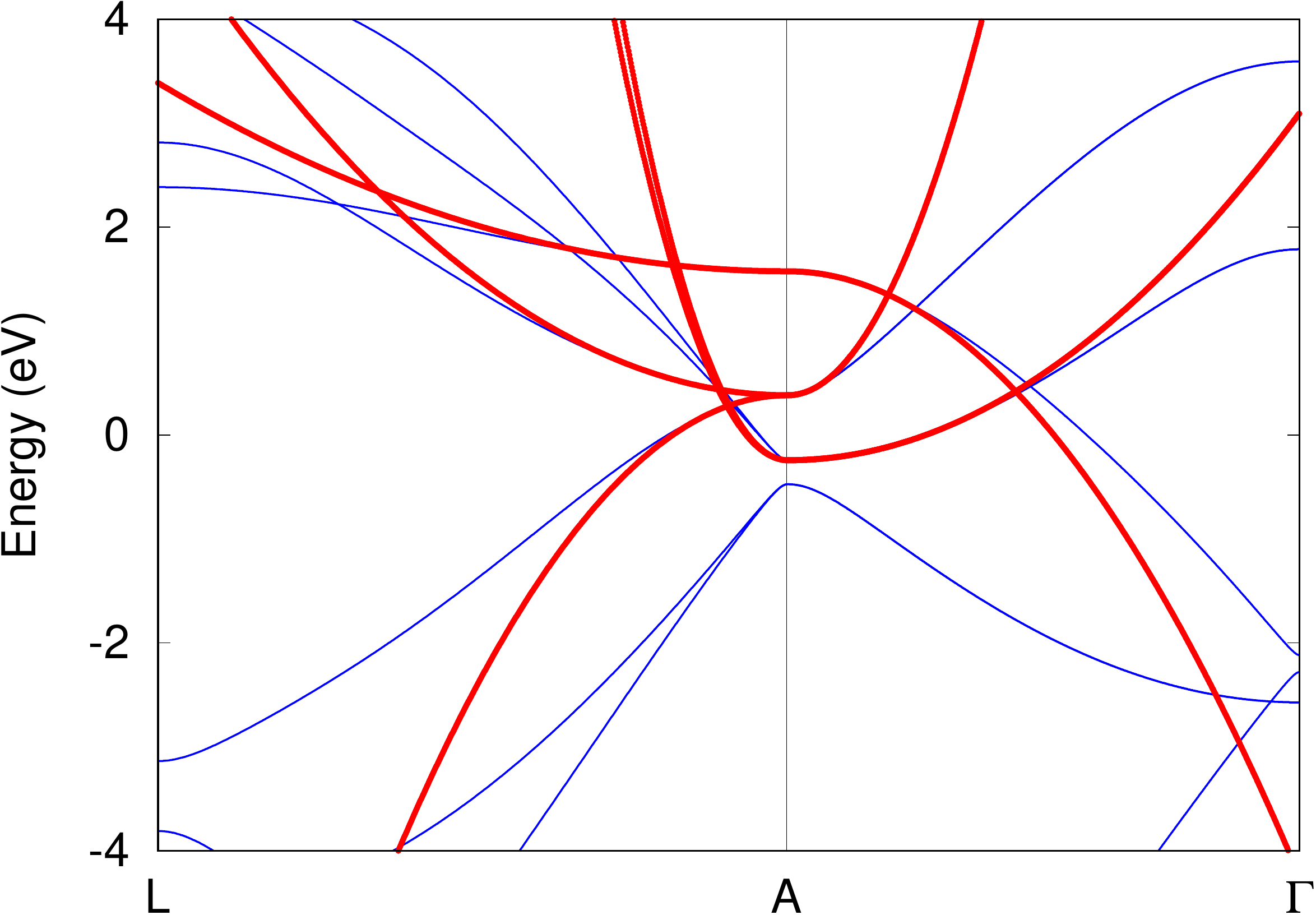}
    \caption{(color online) Comparison between DFT bands and the bands from the $k\cdot p$ model [Eq.(\ref{eqn:kp-A})] at $A$ point of TiB$_2$. Blue lines come from DFT calculation, and red dotted lines come from $k\cdot p$ model. }
    \label{fig:kp-dft-A}
\end{figure}

 The constructed $k\cdot p$ model is given by
\begin{align}
H(k)=\left(\begin{array}{ccccc}
\varepsilon_2(k) & D k_z k_+  & D k_z k_-  & E k_+^2  & -E k_-^2\\
D k_z k_- & \varepsilon_1(k) & C k_-^2 & F k_z k_+ & 0  \\
D k_z k_+ & C k_+^2 & \varepsilon_1(k) & 0 &-F k_z k_-\\
E k_-^2 & F k_z k_- & 0 & \varepsilon_3(k) & G k_+^2 \\
-Ek_+^2 & 0 & -F k_z k_+ & G k_-^2 & \varepsilon_3(k)
\end{array}\right)\label{eqn:kp-A}
\end{align}
where $\varepsilon_i(k)= E_i+ A_i(k_x^2+k_y^2)+ B_i k_z^2$ with $i=1, 2, 3$. The fitted parameters are $E_1=-0.2426\si{\electronvolt}$, $A_1 = 170\si{\electronvolt \angstrom}^2$, $B_1 = 3.53\si{\electronvolt \angstrom}^2$, $E_2 = 1.575\si{\electronvolt}$, $A_2 = 1.27\si{\electronvolt \angstrom}^2$, $B_2 = -6.16\si{\electronvolt \angstrom}^2$, $E_3 = 0.3822\si{\electronvolt}$, $A_3= -2.39\si{\electronvolt \angstrom}^2$
$B_3 = 26.55\si{\electronvolt \angstrom}^2$, $C = 0.47\si{\electronvolt \angstrom}^2$, $D=0.0$, $E=0.0$, $F=0.1\si{\electronvolt \angstrom}^2$, $G=5.62\si{\electronvolt \angstrom}^2$. The fitted band is shown in Fig.\ref{fig:kp-dft-A}.

Particularly, when the effective $k_z=0$, this means that bulk $k_z=0.5$, Eq.(\ref{eqn:kp-A}) decouples into three block diagonal matrices. From Fig.\ref{fig2}, it is shown that along $A-H$, band crossing happens between $d_z^2$ and the combination of $d_{xz}$ and $d_{yz}$. \emph{i.e.} we only have to consider the following block of Eq.(\ref{eqn:kp-A})
\begin{align}
H(k)=\left(\begin{array}{cccc}
 \varepsilon_1(k) & C k_-^2 & 0 & 0  \\
 C k_+^2 & \varepsilon_1(k) & 0 &0\\
 0 & 0 & \varepsilon_3(k) & G k_+^2 \\
 0 & 0 & G k_-^2 & \varepsilon_3(k)
\end{array}\right)\label{eqn:kp-A-block}
\end{align}

\subsection{$k\cdot p$ model for surface states at $\bar{K}$ points}
The little group at $\bar{K}$ point of slab system TiB$_2$ is $C_{3v}$, which has two generators $c_{3z}(-\frac{x}{2}-\frac{\sqrt{3}}{2}y, \frac{\sqrt{3}}{2}x-\frac{y}{2}, z)$ and $\sigma_v(x,-y,z)$. According to the DFT calculations, it was determined that the surface state at $\bar{K}$ belongs to $E$ (see Table 49 of Ref.\cite{Koster1962}) representation of $C_{3v}$. On the basis of complex orbitals, the related representations of its generators are given by
\begin{align}
D(c_{3z})= \left(\begin{array}{cc}
e^{-2i\pi/3} & 0 \\
0 & e^{2i\pi/3}
\end{array}\right), \sigma_v=\left(\begin{array}{cc}
0 & -1 \\
-1 & 0
\end{array}\right)
\end{align}
Considering these symmetries above and the constraint of Eq.(\ref{eqn:symmetry}), a 2-band model up to the second order of k around $\bar{K}$ point for surface states is given by
\begin{align}
H(\bf{k})=\left(\begin{array}{cc}
Bk^2 & Ak_+ +Ck_-^2 \\
Ak_- +Ck_+^2 & Bk^2
\end{array}\right)\label{eqn:sskp}
\end{align}
where $k_{\pm}=k_x\pm ik_y$ and $k^2=k_x^2+k_y^2$. The linear part of Eq.(\ref{eqn:sskp})  leads to massless Dirac dispersion $E(k)=\pm |k|$. The combination of the linear term and the quadratic term leads to 3-fold rotation symmetry of the energy dispersion. Fitting to DFT calculation band structure, We obtain the following parameters for Ti-terminated surface:  $A=0.08\si{\electronvolt \angstrom}, B=-0.46 \si{\electronvolt \angstrom}^2, C= 2.3\si{\electronvolt \angstrom}^2$, and for the B-terminated surface A=1.5\si{\electronvolt \angstrom}, B=0, C=0.

\begin{figure}[!htp]
    \centering
    \includegraphics[width=0.45\textwidth]{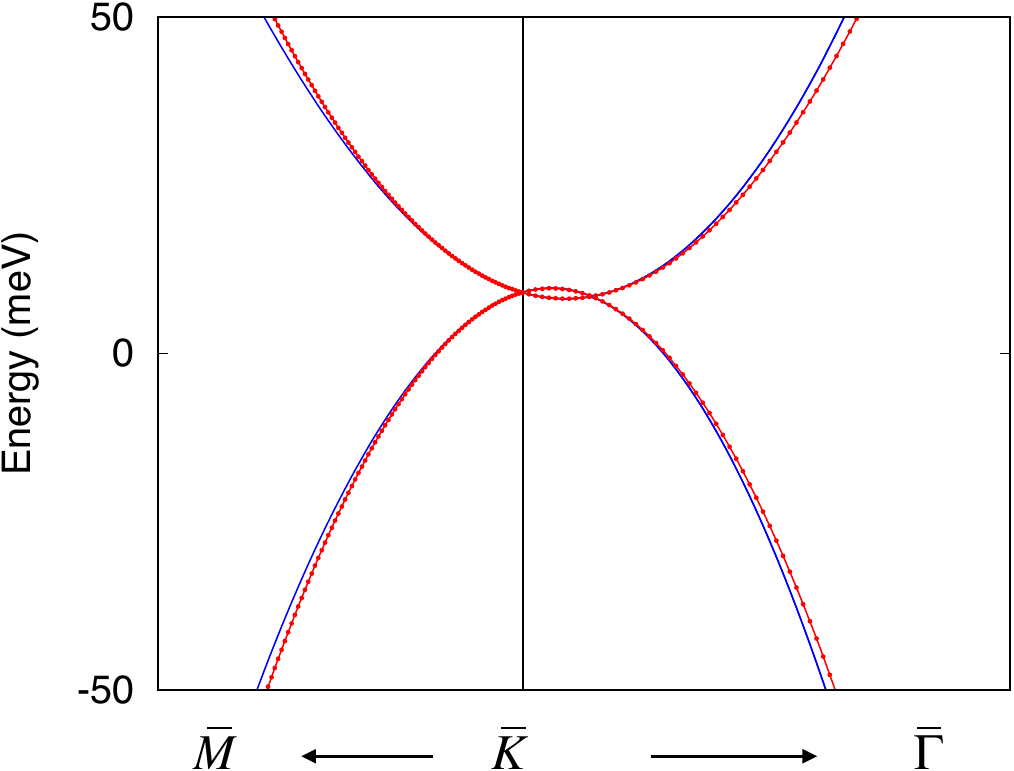}
    \caption{(color online) Comparison between DFT bands and the bands from the $k\cdot p$ model [Eq.(\ref{eqn:sskp})] at $\bar{K}$ point for Ti-terminated surface of TiB$_2$. Blue lines come from DFT calculation, and red dotted lines come from $k\cdot p$ model. }
    \label{fig:kp-dft-K}
\end{figure}

\section{Parities at TRIMs}\label{appendix:toponumber}
The parities of the occupied bands of TiB$_2$ at TRIMs are listed in Table.\ref{tab:parity}. It is noted that there are six occupied bands at $A$, and five occupied bands at other TRIMs. The product of the parities of the occupied bands at $M$ and $L$ are 1, which leads to $\xi_M=1$ and $\xi_L=1$.
\begin{table}[htpb]
\begin{center}
\caption{Parity of the occupied bands of TiB$_2$ at TRIMs, $\Gamma_i$ is in unit of the reciprocal lattice vectors. 'Total' means the product of all parities of occupied bands. $\Gamma_0$, $\Gamma_1(\Gamma_2)$, $\Gamma_3$ and $\Gamma_5$ are the same as the notional $\Gamma$, $M$, $A$ and $L$ respectively.}\label{tab:parity}
\begin{tabular}{c|cccccc|c}
TRIM&	&	&& parity&	& &	Total\\	
\hline
$\Gamma_0$ (0.0, 0.0, 0.0)&	+&	+&	+&	-&	+&	&	-	\\
$\Gamma_1$ (0.5, 0.0, 0.0)&	+&	-&	-&	+&	+&	&	+	\\
$\Gamma_2$ (0.0, 0.5, 0.0)&	+&	-&	-&	+&	+&	&	+	\\
$\Gamma_3$ (0.0, 0.0, 0.5)&	-&	+&	-&	-&	+&	+&	-	\\
$\Gamma_4$ (0.5, 0.5, 0.0)&	+&	-&	-&	+&	+&	&	+	\\
$\Gamma_5$ (0.0, 0.5, 0.5)&	+&	-&	+&	+&	-&	&	+	\\
$\Gamma_6$ (0.5, 0.0, 0.5)&	+&	-&	+&	+&	-&	&	+	\\
$\Gamma_7$ (0.5, 0.5, 0.5)&	+&	-&	+&	+&	-&	&	+	
\end{tabular}
\end{center}
\label{default}
\end{table}%
\clearpage
\bibliography{reference-v2}

\end{document}